\documentclass[%
 reprint,
superscriptaddress,
amsmath,amssymb,
 aps, prl
]{revtex4-1}

\usepackage[utf8]{inputenc}
\usepackage{graphicx, amsmath}
\usepackage{microtype}
\usepackage{dcolumn}
\usepackage{xcolor,soul}
\usepackage{array}
\usepackage{natbib}

\usepackage{siunitx}
\usepackage{bm}
\usepackage[normalem]{ulem}

\definecolor{purp}{rgb}{0.4,0.2,0.8}
\definecolor{gr}{rgb}{0.1,0.6,0.1}

\definecolor{colreva}{rgb}{0,0,0}
\definecolor{colrevb}{rgb}{0,0,0}
\newcommand{\reva}[1]{\textcolor{colreva}{#1}}
\newcommand{\revb}[1]{\textcolor{colrevb}{#1}}

\begin{document}
\title{Bendability parameter for twisted ribbons to describe longitudinal wrinkling and delineate the near-threshold regime}
\author{Madelyn Leembruggen}
\email{(she/her) mleembruggen@g.harvard.edu}
\affiliation{Department of Physics, Harvard University, Cambridge, MA 02138, USA}
\author{Jovana Andrejevic}
\email{jovana@sas.upenn.edu}
\affiliation{Department of Physics, University of Pennsylvania, Philadelphia, Pennsylvania 19104, USA}
\author{Arshad Kudrolli}
\email{akudrolli@clarku.edu}
\affiliation{Department of Physics, Clark University, Worcester, Massachusetts 01610, USA}
\author{Chris H. Rycroft}
\email{chr@math.wisc.edu}
\affiliation{Department of Mathematics, University of Wisconsin--Madison, Madison, WI 53706, USA}
\affiliation{Computational Research Division, Lawrence Berkeley Laboratory, Berkeley, CA 94720, USA}

\begin{abstract}
    We propose a dimensionless bendability parameter, $\epsilon^{-1} = [\left(h/W\right)^2 T^{-1}]^{-1}$ for wrinkling of thin, twisted ribbons with thickness $h$, width $W$, and tensional strain $T$. Bendability permits efficient collapse of data for wrinkle onset, wavelength, critical stress, and residual stress, demonstrating longitudinal wrinkling's primary dependence on this parameter. This new parameter also allows us to distinguish the highly bendable range ($\epsilon^{-1} > 20$) from moderately bendable samples ($\epsilon^{-1} \in (0,20]$). We identify scaling relations to describe longitudinal wrinkles that are valid across our entire set of simulated ribbons. When restricted to the highly bendable regime, simulations confirm theoretical near-threshold (NT) predictions for wrinkle onset and wavelength.
\end{abstract}

\maketitle

Wrinkling of geometrically frustrated sheets is a well-studied subject. Recently, systematic treatments of thin sheet wrinkling examined flat sheets on compressed substrates~\cite{cerda_geometry_2003,chen_surface_2012} and frustrated annuli~\cite{geminard_wrinkle_2004,davidovitch_prototypical_2011,davidovitch_nonperturbative_2012,bella_wrinkles_2014}, which possess tractable axial symmetries. Next thin films were floated on deformable fluid surfaces~\cite{huang_capillary_2007,vella_capillary_2010,holmes_draping_2010,pineirua_capillary_2013,toga_drop_2013,vella_indentation_2015,box_indentation_2017} or adhered to curved planes~\cite{grason_universal_2013,hohlfeld_sheet_2015,davidovitch_geometrically_2019}, further complicating the forces complicit in wrinkle formation. 
Many features of such wrinkled sheets were successfully described using near-threshold (NT) approximations which assume wrinkling amplitude is a small perturbation from the flat state~\cite{davidovitch_prototypical_2011,box_indentation_2017,davidovitch_nonperturbative_2012}. In some cases it was helpful to use a far-from-threshold (FT) expansion which assumes compressive stress in the sheet is alleviated to first order by onset of wrinkles~\cite{cerda_geometry_2003,davidovitch_prototypical_2011,davidovitch_nonperturbative_2012,vella_indentation_2015,davidovitch_geometrically_2019,hohlfeld_sheet_2015,grason_universal_2013,bella_wrinkles_2014}.

A thin ribbon when twisted also develops wrinkles due to geometric frustration. Wrinkled ribbons were first documented in the 1930s~\cite{green_elastic_1937}, but the longitudinally wrinkled phase was not verified numerically until nearly 50 years later~\cite{crispino_stability_1986}. Since then, twisted ribbons have been theoretically analyzed~\cite{coman_asymptotic_2008,chopin_roadmap_2014,pham_dinh_cylindrical_2016}, and their phase space experimentally mapped~\cite{chopin_helicoids_2013}. We previously showed simulations can replicate the morphology and mechanics of ribbons buckled and wrinkled via twisting~\cite{leembruggen_computational_2023} and investigated NT and FT predictions across a broad swath of the parameter space. Despite some of the successes of NT and FT approximations for the twisted ribbon~\cite{coman_asymptotic_2008,chopin_roadmap_2014,chopin_extreme_2019}, the transition between their regimes of validity remains elusive. Exact predictions for onset and wavelength of wrinkling are lacking~\cite{leembruggen_computational_2023}. Simulations offer a grip on ribbons that are neither very thin nor free of compressive stress, with potential to illuminate the murky demarcation of NT and FT predictions.

\begin{figure*}
    \centering
    \includegraphics[width=17.2 cm]{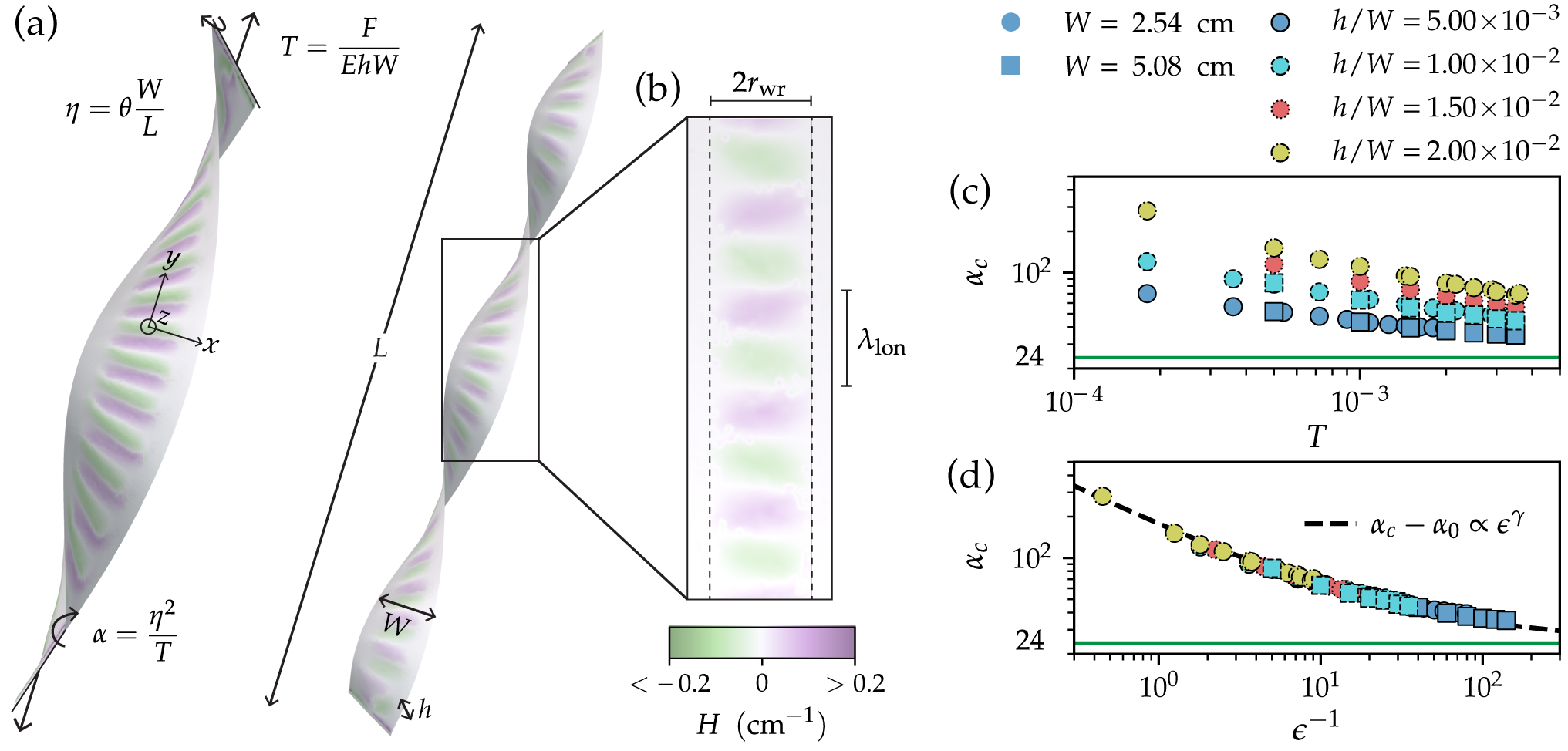}
    \caption{
    (a) Observed snapshots of wrinkled ribbons with width $W = \SI{5.08}{\centi\meter}$ (left) and $\SI{2.54}{\centi\meter}$ (right) at tensions $T$ and twisted by $\eta$ are displayed here, both with thickness $h = \SI{0.254}{\milli\meter}$ and $T = 2.5 \times 10^{-3}$. Overlaid in purple (green) is the mean curvature $H$ of the ribbon, buckled above (below) the plane. (b) The ribbon projected on a flat plane with $H$ rendered according to color map shows the wrinkling is confined to a region of width $2r_\text{wr}$, marked here by vertical dashed lines, and has a wavelength $\lambda_\text{lon}$. (c) The critical confinement, $\alpha_c$, at which wrinkles appear plotted against longitudinally applied tension. An infinitely thin ribbon would develop wrinkles at confinement $\alpha_0 = 24$ (solid green line). Ribbons with finite thickness, however, transition to the wrinkled state at $\alpha_c > \alpha_0$. (d) $\alpha_c$ versus the bendability parameter, defined in Eq.~\eqref{eq:bendability}, with data collapsing onto the line $\alpha_c - \alpha_0 \propto \epsilon^{\gamma}$ with $\gamma = 0.585 \pm 0.004$.}
    \label{fig:onset}
\end{figure*}

\textit{Bendability Parameter---} A thin, twisted ribbon develops longitudinal wrinkles when a longitudinal force $F$ and end-to-end twist $\theta$ are applied, as shown in Fig.~\ref{fig:onset}(a). The wrinkles have wavelength $\lambda_\text{lon}$ and are confined to the ``wrinkled zone'', which is symmetric about the center line with width $2 r_\text{wr}$, labeled in Fig.~\ref{fig:onset}(b).

The two dimensionless parameter groupings useful for analyzing wrinkling in other thin sheets, such as frustrated annuli or floating films, are the confinement (here called $\alpha$) and the bendability ($\epsilon^{-1}$)~\cite{davidovitch_prototypical_2011,davidovitch_nonperturbative_2012,vella_indentation_2015,grason_universal_2013,hohlfeld_sheet_2015,paulsen_curvature-induced_2016,box_indentation_2017,davidovitch_geometrically_2019}. For twisted ribbons, the appropriate confinement parameter was previously identified~\cite{chopin_roadmap_2014}:
\begin{equation}
    \alpha \equiv \frac{\eta^2}{T}
\end{equation}
where $\eta$ represents the geometric strain, and $T$ the tensional strain. $\eta$ and $T$ are themselves dimensionless,
\begin{equation}
    \eta = \theta \frac{W}{L}, \qquad T = \frac{F}{E h W},
\end{equation}
where $E$ is Young's modulus, $h$ is ribbon thickness, $W$ the ribbon width, and $L$ the ribbon length.

On the other hand, bendability for the twisted ribbon, the dimensionless $\epsilon^{-1}$, has not yet been identified in the literature. We propose
\begin{equation}
    \epsilon \equiv \frac{12\left(1-\nu^2\right) B}{WF} = \left(\frac{h}{W}\right)^2 \frac{1}{T}
    \label{eq:bendability}
\end{equation}
to define inverse bendability for twisted ribbons, where $\nu$ is Poisson's ratio and $B$ is the bending modulus.

Previous studies that use bendability, such as ~\cite{davidovitch_prototypical_2011,davidovitch_nonperturbative_2012,vella_indentation_2015,grason_universal_2013,hohlfeld_sheet_2015,paulsen_curvature-induced_2016,box_indentation_2017,davidovitch_geometrically_2019}, focus primarily on the highly bendable limit, such that $\epsilon^{-1} \gg 1$. Many of our ribbon samples fall into a ``moderately'' bendable range with $\epsilon^{-1} \in (0,20]$. We first identify scaling laws for the full range of samples, then examine only the highly bendable sheets ($\epsilon^{-1}>20$) to compare directly to theoretical NT estimates.

\textit{Simulation Details---} Ribbons are simulated by an underlying topology of either $11250$ or $22500$ randomly distributed nodes connected by in-plane springs, and bending between adjacent facets is quadratically penalized~\cite{seung_defects_1988}. Both in-plane stretching and out-of-plane bending models are generalized to the random mesh using modifications adapted from Van Gelder/Lloyd~\cite{van_Gelder_1998,lloyd_identification_2007} and Grinspun~\cite{grinspun_discrete_2003,tamstorf_discrete_2013}, respectively. Full derivation and validation of the model are presented by Leembruggen, {\it et al.}~\cite{leembruggen_computational_2023}.

Interactions between nodes are described by Newton's second law. Since the ribbon is in the quasi-static limit where $F \approx 0$, we use an implicit integration scheme. A typical ribbon simulation requires a wall-clock time of \SI{8}{h}, using 8 threads on a Linux computer with dual \SI{2.40}{GHz} Intel Xeon E5-2630 CPUs.

Six ribbons with $L = \SI{45.7}{\centi\meter}$ were used in this study: two of width $W = \SI{5.08}{\centi\meter}$ with thicknesses $h/W = 5 \times 10^{-3}, 10^{-2}$; and four of width $W = \SI{2.54}{\centi\meter}$ with thicknesses $h/W = 5 \times 10^{-3}, 1 \times 10^{-2}, 1.5 \times 10^{-2}, 2 \times 10^{-2}$. Young's modulus for these ribbons was $E = \SI{3.4}{\giga\pascal}$, and Poisson's ratio was $\nu = 1/3$~\cite{seung_defects_1988,lloyd_identification_2007}. \reva{The short edges of each ribbon were fixed to a rigid rod rotating at constant rotational velocity, $\dot{\theta}$; thus the position of each node on the boundary was imposed at each time step.} Ribbons were additionally held at fixed tensions, $T$, within the longitudinal buckling phase. Ultimately we had 96 samples of longitudinally wrinkled ribbons with varying $\lambda_\text{lon}$. 

\textit{Wrinkling Onset---} Examining the ribbon's stress in the longitudinal ($y$) direction~\cite{chopin_roadmap_2014},
\begin{equation}
    \frac{\sigma^{yy}(x)}{T} = 1 + \frac{\alpha}{2}\left(\left(\frac{x}{W}\right) - \frac{1}{12}\right),
    \label{eq:long_stress}
\end{equation}
identifies $\alpha_0 = 24$ as the confinement at which stress becomes compressive along $x = 0$ (the ribbon's spine). An infinitely thin ribbon, unable to support compressive stress, buckles at $\alpha_0$. But ribbons with thickness support stress, and thus buckle at confinements $\alpha_c > \alpha_0$, as demonstrated across samples in Fig.~\ref{fig:onset}(d). \revb{(Details concerning the appearance of wrinkles, determination of wrinkle onset, and calculation of wavelength are presented by Leembruggen {\it et al.} in Fig. 4 of Ref.~\cite{leembruggen_computational_2023}.)} Previous estimates of this finite thickness correction based on experiments followed the form $\eta_c = \eta_0 + C_\text{lon} \frac{h}{W}$ where $\eta_0 = \sqrt{24T}$ coincides with $\alpha_0 = 24$~\cite{chopin_helicoids_2013}. Written in terms of proposed bendability, this translates to a correction on the order of $\alpha_c - \alpha_0 \propto \epsilon$. However, as plotted in Fig.~\ref{fig:onset}(e), we observe
\begin{equation}
    \alpha_c - \alpha_0 \propto \epsilon^{\gamma},
    \label{eq:alpha_scaling}
\end{equation}
with $\gamma = 0.585 \pm 0.004$. As shown by the data collapse in Fig.~\ref{fig:onset}(e) the tension dependence and geometric parameters are captured by the bendability parameter introduced in Eq.~(\ref{eq:bendability}). The previously proposed correction of $\mathcal{O}(\epsilon)$ overcompensated for the thickness of the ribbon.
Chopin {\it et al.}\@ posit that, using the NT approximation, $\left(\alpha_c - \alpha_0\right) \sim (h/W)T^{-1/2}$~\cite{chopin_roadmap_2014}. Recast using $\epsilon^{-1}$, the equivalent prediction is $\alpha_c - \alpha_0 \propto \epsilon^{\gamma}$ with $\gamma=1/2$. 

\begin{figure}
    \centering
    \includegraphics[width=\columnwidth]{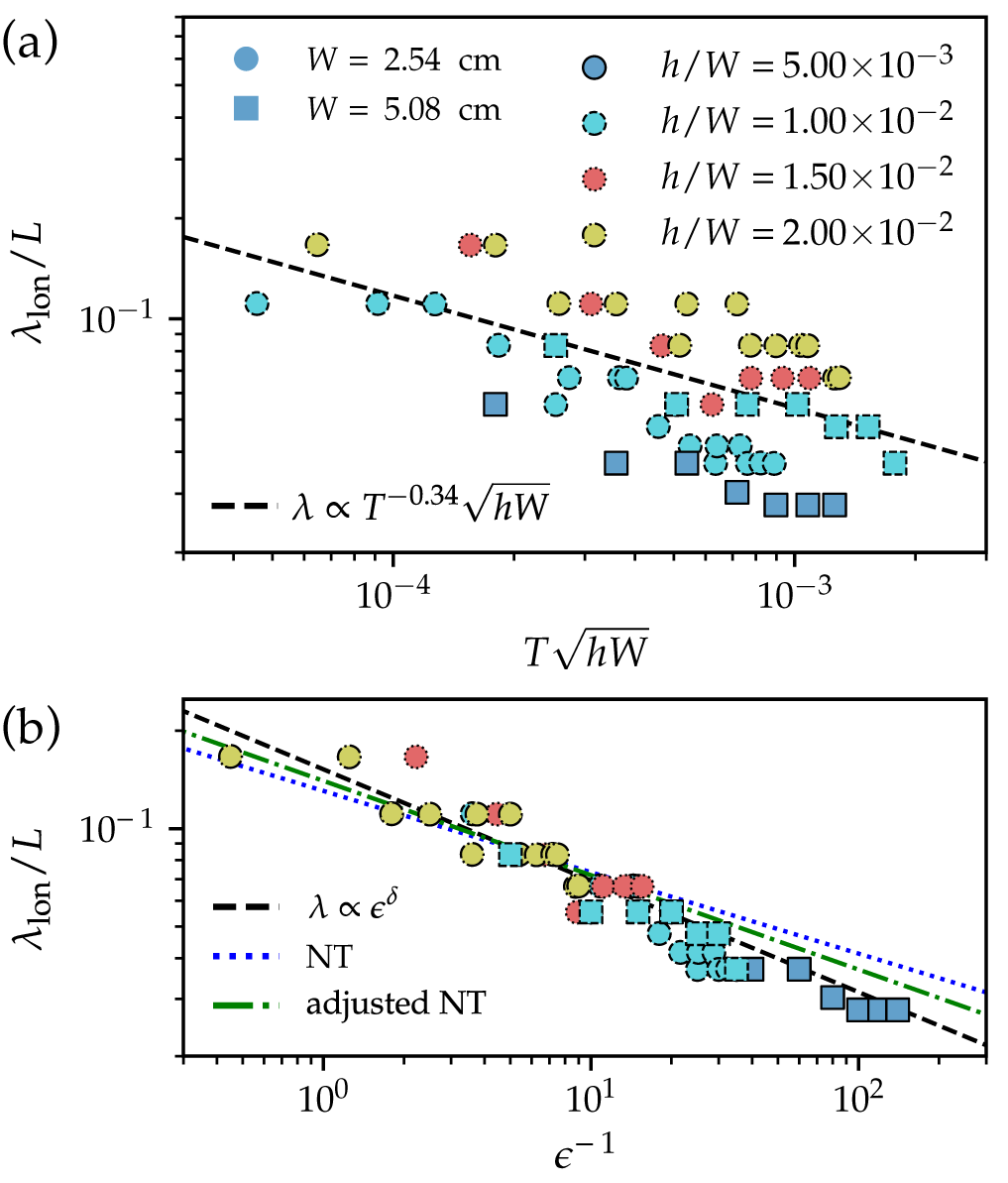}
    \caption{(a) \reva{Wavelength per unit length as a function of ribbon dimensions and applied tension, compared with Chopin and Kudrolli's experimental normalization (dashed black line)~\cite{chopin_helicoids_2013}}. Plotted with this normalization, data are broadly distributed. (b) Plotted against bendability, wavelengths collapse more closely. In terms of $\epsilon^{-1}$, theoretical NT analyses~\cite{coman_asymptotic_2008, chopin_roadmap_2014} predict $\delta = 1/4$, plotted by the blue, dotted line. Adjusting the NT approximation by substituting $\gamma$ observed in Fig.~\ref{fig:onset}(e), such that $\delta = \gamma/2 = 0.293 \pm 0.002$ is shown here by the green, dash-dotted line. Note that points with $\epsilon^{-1}>20$ roughly follow the slope of the NT prediction, but diverge from this line when $\epsilon^{-1}<20$.}
    \label{fig:wavelength}
\end{figure}

Wavelength at onset of longitudinal wrinkling is constant in $\alpha$, so we wait until the wrinkle pattern has appreciable amplitude before extracting $\lambda_\text{lon}$ to make the measurement more precise~\cite{leembruggen_computational_2023}. NT approximations have been used to estimate wavelength at onset~\cite{coman_asymptotic_2008,chopin_roadmap_2014}: $\lambda \propto r_\text{wr} \propto \sqrt{h/W} T^{-1/4}$, and were noted to describe experimentally measured wavelengths~\cite{chopin_helicoids_2013}.

In Fig.~\ref{fig:wavelength}(a) we plot wavelength per unit length (i.e. the inverse of preferred wave number) using the same normalization as Chopin and Kudrolli's experiment~\cite{chopin_helicoids_2013} \reva{which differs from the NT prediction by a factor of $W$ but has the same dependence on $T$}. Also plotted in this figure is a best fit as a function of $T$, represented by the black, dashed line. Vertical spread in these data suggests the dependence on $h$ and $W$ is not captured by $\sqrt{hW}$ normalization. However, plotting wavelengths against the bendability, in Fig.~\ref{fig:wavelength}(b), the values collapse, scaling as $\lambda_\text{lon}/L \propto \epsilon^{\delta}$ with $\delta = 0.342 \pm 0.019$, shown as a black
, dashed line.

Recast in terms of bendability, the theoretical NT prediction for wavelength is $\lambda_\text{lon}/L \propto (\alpha_c - \alpha_0)^{1/2} \propto \epsilon^{\delta}$, with $\delta = \gamma/2 =1/4$ in this case. Alternatively we could assume adjusting $\gamma$ propagates to other NT predictions. In which case, $\delta = \gamma/2 = 0.293 \pm 0.002$, using Eq.~\eqref{eq:alpha_scaling}. In Fig.~\ref{fig:wavelength}(b), we plot the theoretical NT $\delta = 1/4$ in a blue, dotted line, and adjusted NT $\delta = 0.293$ in a green, dash-dotted line.

All in all, scaling $\lambda_\text{lon}$ in terms of $\epsilon^{-1}$ is a vast improvement over past investigations of wavelength versus tension. Regardless of the collapse's precise slope, the points are clustered more closely than in previous studies~\cite{leembruggen_computational_2023}, further affirming $\epsilon$'s usefulness.

\begin{figure}
    \centering
    \includegraphics[width=\columnwidth]{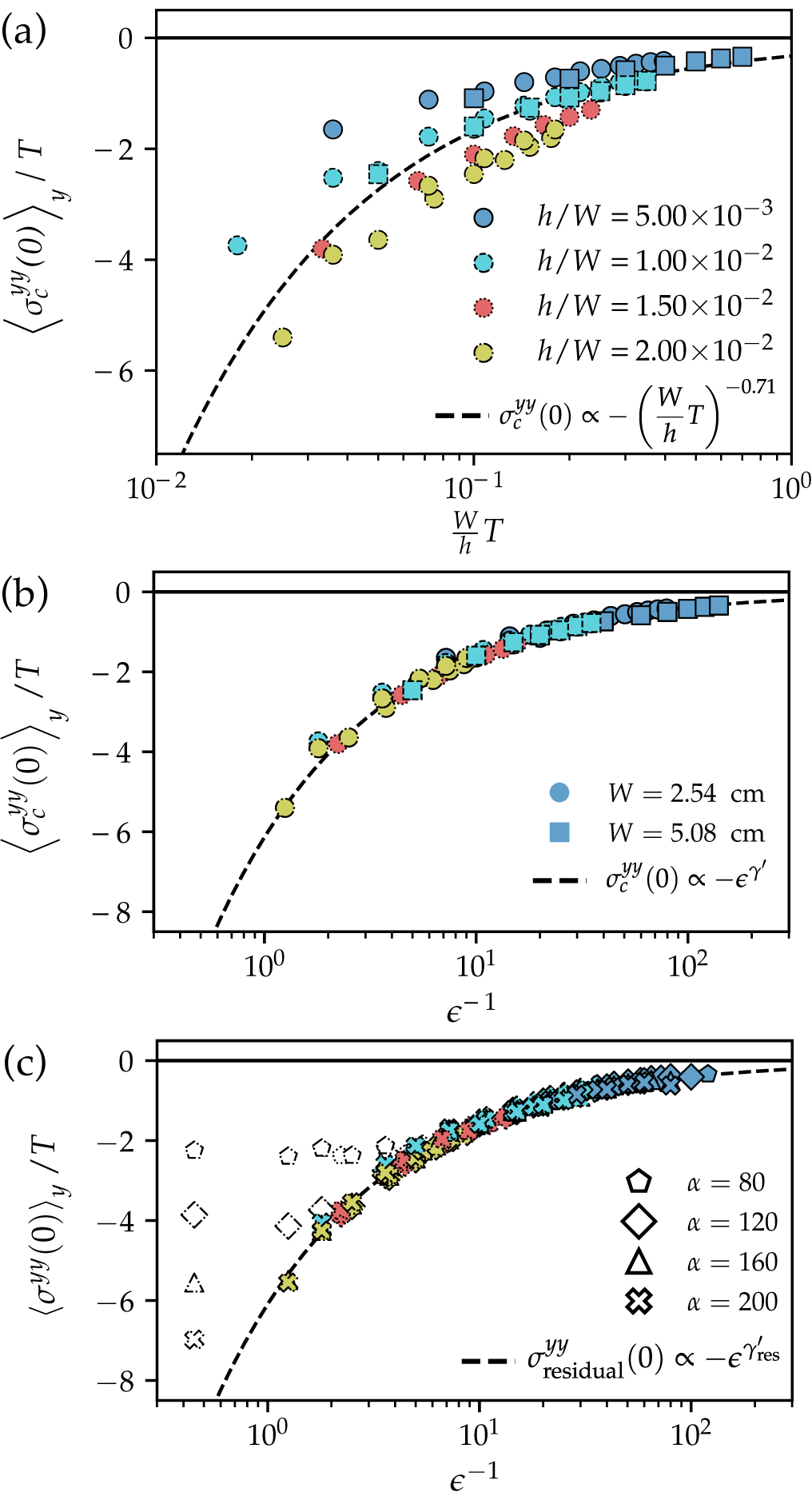}
    \caption{(a) The critical stress along the ribbon center at onset of wrinkling plotted against $\frac{W}{h}T$ results in incomplete collapse of the data. (b) When written as a function of bendability, stress at critical confinement follows the curve $\sigma^{yy}_c\left(0\right) \propto -\epsilon^{\gamma'}$, $\gamma' = 0.599 \pm 0.006$. This relationship follows from Eqs.~\eqref{eq:long_stress} \& \eqref{eq:sigma_scaling}. (c) The stress of each ribbon sample at four different confinements, $\alpha$, resulting in $384$ points. Open markers indicate the stress in a ribbon before it has buckled, whereas filled markers indicate the stress post-buckling. Marker shape refers to the confinement, $\alpha$, at which the stress was sampled. Post-buckling, the ribbons support residual stress. Residual stress saturates according to ribbon bendability, and is independent of $\alpha$. Post-buckling residual stress follows the relationship $\sigma^{yy}_\text{residual}\left(0\right) \propto -\epsilon^{\gamma'_\text{res}}$, $\gamma_\text{res}' = 0.591 \pm 0.003$, similar to the scaling of $\sigma^{yy}_\text{c}$ in (b).}
    \label{fig:stress}
\end{figure}

\textit{Residual Stress---} Throughout the simulation we extract the stress tensor via the mesh's deformation gradient tensor~\cite{leembruggen_computational_2023}. Longitudinal slices are taken along the middle third of the ribbon, then averaged along the $y$ direction to obtain $\left<\sigma^{yy}\left(x\right)\right>_y$. Instinctually, one might expect critical buckling stress $\left<\sigma^{yy}_c\left(0\right)\right>_y$ is proportional to cross-sectional area ($hW/W^2 = h/W$ in dimensionless units). In Fig.~\ref{fig:stress}(a) we plot critical stress at $x=0$ against this candidate scaling factor. This simple estimate, however, results in incomplete collapse of the data.

Turning back to our stress equation, Eq.~\eqref{eq:long_stress}, along $x = 0$ we expect the stress at $\alpha>\alpha_0$ to be
\begin{equation}
  \frac{\sigma^{yy}}{T} = 1-\frac{\alpha}{24} = 1 - \frac{\alpha_0}{24} - \frac{\alpha-\alpha_0}{24} = - \frac{\alpha - \alpha_0}{24}.
\end{equation}
Since we observe the scaling for confinement in Eq.~\eqref{eq:alpha_scaling}, critical stress should go as
\begin{equation}
    \frac{\sigma^{yy}_c}{T} = -\frac{\alpha_c - \alpha_0}{24} \propto -\epsilon^{\gamma}.
    \label{eq:sigma_scaling}
\end{equation}
This is consistent with our measurements, plotted in Fig.~\ref{fig:stress}(b), which find $\sigma_c^{yy}/T \propto -\epsilon^{\gamma'}$, with the measured $\gamma' = 0.599 \pm 0.006$.

It is not clear from analytical studies whether the ribbon's residual, post-buckling, compressive stress should scale the same as its critical stress, especially as $\alpha>\alpha_c$ changes~\cite{qiu_morphology_2017,davidovitch_geometrically_2019}. However, as shown in Fig.~\ref{fig:stress}(c), stress saturates at values very close to critical stress; we measure $\gamma_\text{res}' = 0.591 \pm 0.003$.

\textit{The Boundary of Near Threshold---} It is expected that NT relations are valid only for highly bendable sheets $(h/W)^2 \ll T$, or equivalently, $\epsilon^{-1} \gg 1$. Although there are no sharp transitions in the data of Figs.~\ref{fig:onset} \& \ref{fig:wavelength} as a function of $\epsilon^{-1}$, we could exclude points with small bendability to better understand how confinement ($\alpha_c - \alpha_0$), wavelength ($\lambda_\text{lon}$), and critical stress ($\langle \sigma^{yy}_c(0)\rangle_y/T$) scale in the highly bendable limit.
\begin{figure}
    \centering
    \includegraphics[width=\columnwidth]{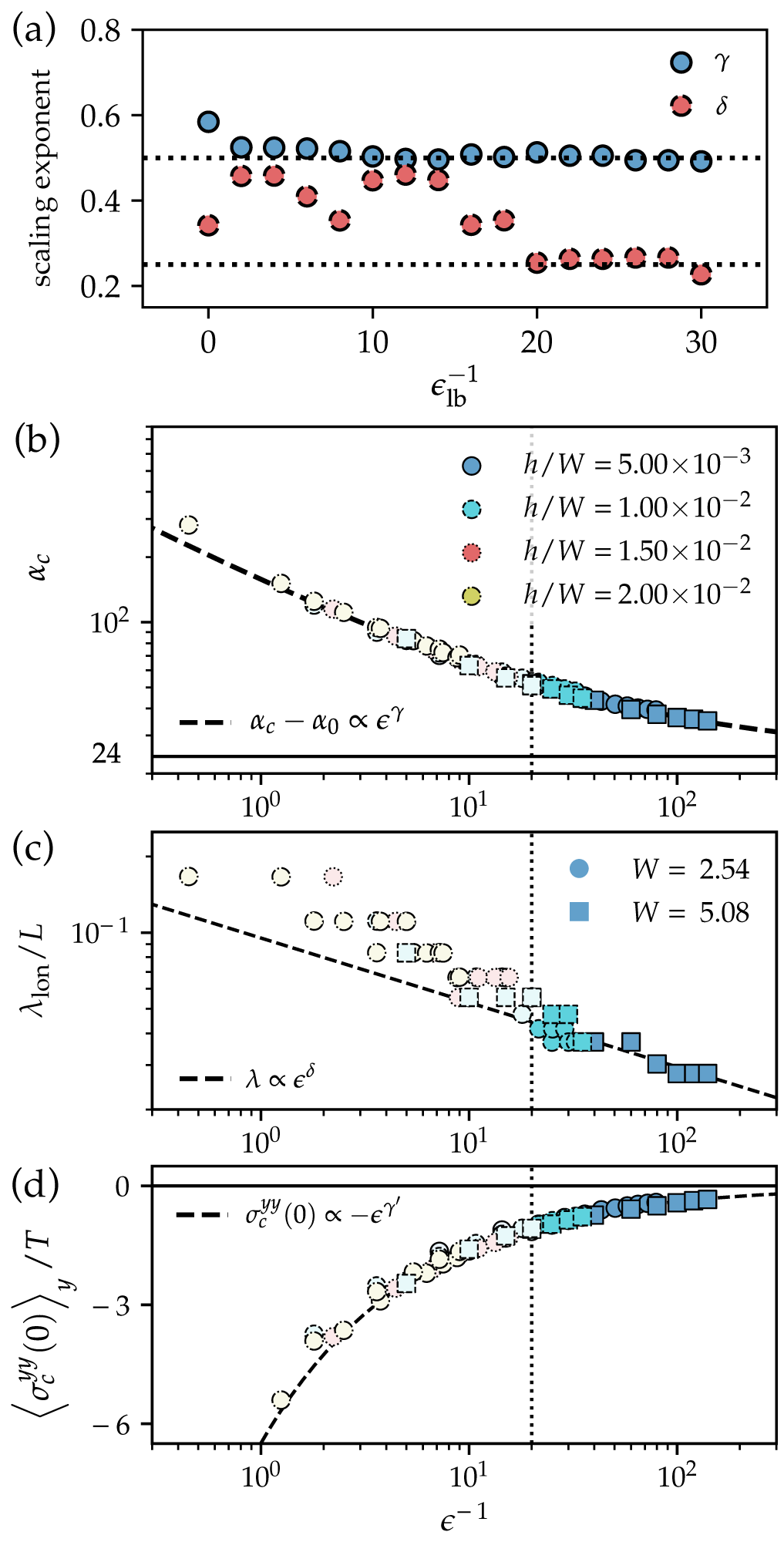}
    \caption{(a) Various thresholds are placed on the lower bound of $\epsilon^{-1}$. The confinement scaling exponent, $\gamma$ (upper, blue circles, solid border) remains largely unchanged as lower bound changes. The wavelength exponent, $\delta$ (lower, red circles, dashed border), is more sensitive to the threshold of $\epsilon^{-1}$; when the threshold reaches $\epsilon^{-1}_\text{lb} \approx 20$, $\delta \approx 0.25$. Horizontal lines are plotted at $0.25$ and $0.5$ to guide the eye. (b--d) Replications of Figs.~\ref{fig:onset}(d), \ref{fig:wavelength}(b) and \ref{fig:stress}(b) where the scaling exponent (dashed, black line) is fitted only using data with $\epsilon^{-1}>20$. Darker points, right of the vertical dotted lines, are included in the scaling fit; desaturated points to the left are not. With this threshold imposed, $\gamma = 0.513 \pm 0.001$ and $\delta = 0.255 \pm 0.044$ (compared to NT predictions of $\gamma = 0.50$ and $\delta = 0.25$). The critical stress scaling remains essentially unchanged, with $\gamma' = 0.608 \pm 0.029$.}
    \label{fig:exp_conv}
\end{figure}

In Fig.~\ref{fig:exp_conv}(a) we plot the value of scaling exponents for confinement ($\gamma$) and wavelength ($\delta$) as a function of the lower threshold, $\epsilon^{-1}_\text{lb}$. As $\epsilon^{-1}_\text{lb}$ increases, further restricting the data to a highly bendable range, $\gamma$ changes little, hovering around $0.5$. On the other hand, $\delta$ is quite sensitive to $\epsilon^{-1}_\text{lb}$, but converges to $\approx 0.25$ when $\epsilon^{-1}_\text{lb}>20$. Therefore, when constrained by a highly bendable limit, our simulations agree entirely with NT predictions that $\left(\alpha_c - \alpha_0\right)_\text{NT} \propto \epsilon^{1/2}$ and $\lambda_\text{lon}^\text{NT} \propto \epsilon^{1/4}$.

Figures \ref{fig:exp_conv}(b--d) recreate Figs.~\ref{fig:onset}(d), \ref{fig:wavelength}(b), \& \ref{fig:stress}(b) respectively, this time restricting the data such that only highly bendable ribbons are included in the scaling exponent fit. Excluded data are desaturated, left of the vertical dotted lines, and included data are fully saturated to the right of the cutoff line. With this threshold imposed, the confinement in Fig.~\ref{fig:exp_conv}(b) has $\gamma = 0.513 \pm 0.001$;  wavelength in Fig.~\ref{fig:exp_conv}(c) has $\delta = 0.255 \pm 0.044$; and critical stress in Fig.~\ref{fig:exp_conv}(d) has $\gamma' = 0.608 \pm 0.029$. Interestingly, critical stress remains largely unchanged when thresholded, and is inconsistent with the stress (Eq.~\eqref{eq:sigma_scaling}) expected using the thresholded confinement scaling.

\textit{Discussion---}Inspired by studies of complementary wrinkled systems, we introduced a dimensionless parameter for bendability of a twisted ribbon $\epsilon^{-1}$. Specifically $\epsilon^{-1}$ incorporates finite thickness, and extends predictions concerning infinitely thin ribbons. Then, we demonstrated $\epsilon^{-1}$ successfully describes onset of wrinkling, wavelength, and even stress at and after wrinkle onset. By rewriting NT and FT predictions~\cite{chopin_roadmap_2014} in terms of $\epsilon^{-1}$, we show onset of wrinkling scales somewhere between NT and FT predictions of~\cite{chopin_roadmap_2014}, but more closely to NT results. When data are restricted to a highly bendable region ($\epsilon^{-1} > 20$), they exactly match NT predictions. \revb{With our simulation framework, additional simulations of even more bendable ribbons could readily be performed to further probe the NT theory.} Perhaps most successfully, $\epsilon^{-1}$ allows investigation of ribbons in the ``moderately'' bendable ($\epsilon^{-1} \in (0,20]$) limit, which are less bendable than theoretically tractable ribbons. Including this broader set of bendabilities enabled determination of scaling laws are valid for a wide range of ribbons.

Further, analyzing simulated ribbons allowed us to extract stress throughout the twist, which is prohibitively difficult to measure in experiments. Thus, the simulations have revealed stress in finite-thickness ribbons is not completely alleviated by buckling, as is assumed in FT analysis, and the compressive stress they support saturates at critical buckling stress. Our simulations also demonstrate critical stress, and thus saturated stress, depends only on ribbon bendability. \revb{Other quantities, such as suppression of wrinkles near the boundaries, and growth of wrinkle amplitude can also likely be scaled according to $\epsilon^{-1}$~\cite{leembruggen_computational_2023}; additional analytical and experimental studies should be performed to verify and identify these relationships.} We anticipate these facts uncovered by our simulations, and introduction of a bendability parameter for twisted sheets, will prove useful in further theoretical development of scaling laws.

\acknowledgments 
ML was supported by the Ford Foundation Predoctoral Fellowship and the National Science Foundation Graduate Research Fellowship Program under grant no. DGE-1745303. AK was supported by National Science Foundation grant DMR-2005090. CHR was partially supported by the Applied Mathematics Program of the U.S.\@ DOE Office of Science Advanced Scientific Computing Research under contract number DE-AC02-05CH11231.

\bibliographystyle{unsrt}
\bibliography{ref}
\end{document}